\documentclass[12pt,preprint]{aastex}
\usepackage{psfig,lscape,natbib}
\def\lap{\lower.5ex\hbox{$\; \buildrel < \over \sim \;$}}
\def\gap{\lower.5ex\hbox{$\; \buildrel > \over \sim \;$}}

\def\ergcm2s{${\rm erg\ cm^{-2}\ s^{-1}}$}

\def\ergscm2s{${\rm erg\ cm^{-2}\  s^{-1}}$}

\def\cm-2{${\rm cm^{-2}}$}

\begin{document}

\title{A Potential Supernova Remnant/X-ray Binary Association in M31}

\author{Benjamin~F.~Williams\altaffilmark{1}, Robin~Barnard\altaffilmark{2}, Michael~R.~Garcia\altaffilmark{1}, U.~Kolb\altaffilmark{2}, J.~P.~Osborne\altaffilmark{3}, and Albert~K.~H.~Kong\altaffilmark{4}} %, and Albert K. H. Kong\altaffilmark{1}} 
\altaffiltext{1}{Harvard-Smithsonian
Center for Astrophysics, 60 Garden Street, Cambridge, MA 02138;
williams@head.cfa.harvard.edu; garcia@head.cfa.harvard.edu}
\altaffiltext{2}{The Department of Physics and Astronomy, The Open
University, Walton Hall, Milton Keynes, MK7 6BT, U.K.;
r.barnard@open.ac.uk; u.c.kolb@open.ac.uk}
\altaffiltext{3}{The University of Leicester, University Road, Leicester, LE1 7RH, U. K.;  julo@star.le.ac.uk}
\altaffiltext{4}{MIT Kavli Institute for Astrophysics and Space
Research, 77 Massachusetts Avenue, Cambridge, MA 02139; akong@space.mit.edu}


\begin{abstract}

The well-studied X-ray/Optical/Radio supernova remnant DDB 1-15
(CXOM31~J004327.8+411829; r3-63) in M31 has been investigated with
archival {\it XMM-Newton} and {\it Chandra} observations.  The timing
data from {\it XMM-Newton} reveals a power density spectrum (PDS)
characteristic of accreting compact objects in X-ray binaries (XRBs).
The PDS shows features typical of Roche lobe overflow accretion,
hinting that the XRB is low-mass.  The {\it Chandra} observations
resolve the SNR into a shell and show a variable count rate at the
94\% confidence level in the northwest quadrant.  Together, these {\it
XMM-Newton} and {\it Chandra} data suggest that there is an XRB in the
SNR r3-63 and that the XRB is located in the northwestern portion of
the SNR.  The currently-available X-ray and optical data show no
evidence that the XRB is high-mass.  If the XRB is low-mass, r3-63
would be the first SNR found to contain a low-mass X-ray binary.

\end{abstract}

\section{Introduction}

Neutron stars and black holes are thought to form in core-collapse
supernova (SN) explosions.  In recent years, several neutron stars
have been found as X-ray point sources within supernova remnants (SNR)
by {\it Chandra} (e.g., G292.0+1.8 \citealp{hughes2001}; G0.9+0.1
\citealp{gaensler2001}; CTA 1 \citealp{halpern2004}), but none of
these compact objects has been in a binary system.  Only a single
supernova remnant/X-ray binary (SNR/XRB) association is known in the
Galaxy, and that is the association between the high-mass X-ray binary
(HMXB) SS~433 and the SNR W50 \citep{geldzahler1980}.

Some possible extragalactic SNR/XRB associations have been discovered
in recent years.  \citet{roberts2003} inferred a black hole XRB in the
ultra-luminous SNR MF16 in the galaxy NGC~6946, and \citet{chu2000}
identified a possible SNR/XRB pair in the LMC (RX J050736-6847.8).
Additionally, \citet{coe2000} suggested XTE~J0111.2-7317 in the SMC as
a possible Be XRB in a SNR.  None of these systems contain a low-mass
X-ray binary (LMXB).  The reason that there are so few known SNR/XRB
associations and no known SNR association with a LMXB is unclear.

Finding SNR/XRB associations requires the ability to reliably classify
X-ray sources.  The classification of LMXBs can be achieved through
analyses of their variability properties, and SNRs can be recognized
by their appearance in resolved images and their soft X-ray spectra.

LMXBs are variable in X-rays.  The properties of their
variability depend more on the accretion rate than on the nature of
the primary. Van der Klis (1994) showed that the power density spectra
(PDSs) of neutron star and black hole LMXBs are remarkably similar at
low accretion rates; such PDSs are well-described by a broken power
law, with the spectral index changing from $\sim$0 to $\sim$1 at a
break frequency in the range 0.01--1 Hz.  An X-ray source that
exhibits such a low accretion rate PDS (Type A; \citealp{bko04})
cannot be an active galactic nucleus (AGN); PDSs of AGNs break at a
frequency that is several decades lower (see e.g.,
\citealp{utt02}). Therefore any source with a low accretion rate PDS
is almost certainly a disk-accreting system, and the most likely
mechanism is via Roche lobe overflow of the secondary, typical of LMXB
systems.

The large effective area of {\it XMM-Newton} has allowed for the first
time detailed analysis of PDSs from X-ray sources outside the Milky
Way and Magellanic Clouds \citep{bko03,bok03,bko04,bokh04}, providing
the data necessary to discover low accretion rate sources in M31.
Because such sources are likely LMXBs, finding one inside of a SNR is
of great interest to studies of the connection between SNe and
LMXBs.

{\it Chandra} and {\it XMM-Newton} data of M31 taken over the past several
years have included the SNR DDB 1-15 \citep{dodorico1980,braun1990}.
This SNR, known in the {\it Chandra} literature as r3-63 or
CXOM31~J004327.8+411829 \citep{kong2002acis}, was the first SNR in M31
ever resolved in X-rays.  The X-ray size and spectrum helped to
constrain the age, temperature, and surrounding interstellar medium
density (\citealp{kong2002snr}; hereafter K02).

Since its first resolved X-ray detection, there have been four {\it
XMM-Newton} observations and four additional resolved {\it Chandra}
observations of r3-63.  In this paper, we discuss our analysis of
these data which provides strong evidence of an associated X-ray
binary in r3-63.  The association would be the first SNR/XRB found in
M31, one of only a handful known, and possibly the first discovery of
a SNR associated with a LMXB.  Section 2 describes the details of
the observations and reductions of the data sets used.  Section 3
gives the results from those analyses, and section~4 discusses the
classification of the XRB system.  Finally, our conclusions are
summarized in section~5.

\section{Data}

\subsection{{\it XMM-Newton}}

We present data from a $\sim$60 ks observation taken 2002 January and
from three shorter observations spanning 2004 July 16--19. Details of
these observations are provided in Table~\ref{data}.  Several
intervals in the 2004 observations were severely contaminated by
flares in the particle background; filtering out these events leaves
$\sim$40 ks of good data from 2004.

This analysis used data from the three co-aligned EPIC instruments:
the two MOS detectors (MOS1 and MOS2, \citealp{turn01}) and the pn
detector \citep{stru01}; each instrument has a co-aligned
$30'\times30'$ field of view.  We obtained lightcurves and spectra
from each observation using the {\it XMM-Newton} science analysis
software (SAS) version 6.0.0. We used a circular extraction region
with 30$''$ radius centered on r3-63; for the background, we used an
equivalent source-free region on the same CCD at a similar offset from
the aim-point.  We detected 2200 counts and 1300 counts from r3-63 in
the 2002 and 2004 data, respectively.  We created source and
background lightcurves from each EPIC instrument in the 0.3--2.0 keV
band with 2.6 s binning.  Finally, we obtained source and background
spectra from the EPIC-pn in the range 0.3--10 keV. Corresponding
response matrices and ancillary response files were then created.  The
source was not detected at energies $>$3 keV.

The data products were analyzed with FTOOLS version 5.3.1 and XSPEC
11.3.1. Combined EPIC, background-subtracted lightcurves of r3-63 were
produced with {\em lcmath}.

\subsection{{\it Chandra}}\label{datsec}

We found five observations in the {\it Chandra} archive which have
r3-63 near enough to the optical axis to be resolved.  The $>$100
other observations of M31 in the {\it Chandra} data archive were not
of use for localizing a point source within the SNR because they did
not have the SNR close enough to the optical axis to reliably resolve
the shell.  The observation identification (OBSID) numbers, dates,
optical axis coordinates, roll angles, exposure times, and off-axis
angles to r3-63 for these 5 resolved ACIS-I observations are provided
in Table~\ref{data}.

The event lists of the observations were aligned and combined.  The
alignment was performed using the CIAO script {\it
align\_evt}\footnote{http://cxc.harvard.edu/cal/ASPECT/align\_evt/help.html},
and the combination was performed with the script {\it
merge\_all}\footnote{http://cxc.harvard.edu/ciao/download/scripts/merge\_all.tar}.
The alignments had rms errors of $\sim$0.2$''$.

We created X-ray three color images of the combined data, totaling 25
ks of exposure time.  We used the energy bands 0.3--0.8 keV (red),
0.8--1.2 keV (green), and 1.2--7 keV (blue), producing an image in
each band with a pixel scale of 0.246$''$ pixel$^{-1}$.  These images
were then smoothed with a Gaussian of $\sigma$=0.58$''$ and recombined
into the color images shown in Figure~\ref{im}. In these images, the
SNR has a diameter of $\sim$11$''$ and 164 background-subtracted
counts for a mean surface brightness of $\sim$4.5$\times$10$^{-16}$
erg cm$^{-2}$ s$^{-1}$ arcsec$^{-2}$ (0.3--7 keV).

The left panel shows the resulting image after the soft-band image was
multiplied by 1.4 to correct for the $\sim$40\% loss in sensitivity of
{\it Chandra} at 0.5 keV observed between launch and
2002\footnote{http://cxc.harvard.edu/cal/Acis/Cal\_prods/qeDeg/index.html}.
This correction factor did not qualitatively change the image.  The
left panel of Figure~\ref{im} shows our 25 ks image of r3-63 without
correcting for the {\it Chandra} response.  Therefore it can be
directly compared to the single-exposure (5 ks) image in K02 which
also shows the {\it Chandra} detected counts without calibrating each
band for the {\it Chandra} response.  The difference between the left
panel of Figure~\ref{im} and Figure 1 of K02 is that our figure
contains 25 ks of exposure and that of K02 contains only 5 ks of
exposure.  The green color is due to the high sensitivity of {\it
Chandra} at 1 keV.

The right panel shows the resulting image after normalizing each band
by the effective exposure in that bandpass.  We created the exposure
maps for each bandpass and normalized the images using the CIAO script
{\it merge\_all}.  The exposure maps correct for the lower sensitivity
of {\it Chandra} at soft energies as well as the effects of the
optical blocking filer (OBF)
contamination\footnote{http://cxc.harvard.edu/cal/Acis/Cal\_prods/qeDeg/index.html},
which has been building up since launch, decreasing the sensitivity at
energies below 1 keV.  Therefore this image is a better representation
of the ``true'' X-ray color of the SNR.

We created lightcurves for 3$''$ aperture radii centered on 20000
randomly chosen positions in the SNR.  These were measured with the
CIAO task {\it lightcurve} using the counts from each of the five 5 ks
observations to measure the corresponding errors from the merged event
list.  To be sure our errors were not underestimated because of the
small number of counts ($\lap$20 ct bin$^{-1}$) in these lightcurves,
we applied Poisson errors for the count rates.  The use of Poisson
errors, which are systematically larger than Gaussian ($\sqrt{N}$)
errors, makes any detection of deviations from a constant count rate
more reliable.

We also created a lightcurve for the SNR as a whole.  This lightcurve
had $>$20 ct bin$^{-1}$, allowing us to use Gaussian errors.  

All of the lightcurves were fit to a constant count rate that produced
the minimum value of $\chi^2$, in order to test for variability across
observations. We also ran Monte Carlo simulations to test the
reliability of the $\chi^2$ statistics when applied to these data.

Finally, we obtained the deepest {\it Chandra} ACIS-S observation
(Observation ID 1575; 37.7 ks) available in the archive.  Source r3-63
was not resolved in these data, but they contained $\sim$270 counts
from the SNR.  We extracted the spectrum using the CIAO script {\it
psextract}, and we fit the spectrum with several model combinations
using CIAO 3.2/Sherpa.  Results are discussed as in \S 3.1.

\section{Results}

\subsection{\it XMM-Newton}

We present the 0.3--2.0 keV combined EPIC lightcurves of r3-63 from
the 2002 and 2004 {\it XMM-Newton} observations in Fig.~\ref{lcs}. The
lightcurves are averaged over 200 s bins.  The lightcurves of each
observation are featureless but exhibit variability at high
confidence.  For example the lightcurve from the 2002 observation has
$\chi^2/\nu = 517/304$ when fitted to a constant count rate.

We accumulated PDSs for the 2002 lightcurve and the 2004 lightcurve
from observation xmm2 (see Table~\ref{data}).  The other 2004
lightcurves were not further analyzed due to high background
variability, so that the 2004 PDS contained 700 counts.  The PDSs were
integrated over many 333 s intervals with 5.2 s bins (64 bins per
interval); the PDSs were Leahy normalized, so that Poisson noise had a
power of 2. The 2002 data were averaged over 191 intervals and the
2004 data were averaged over 60. The resulting PDSs are presented in
Figure~\ref{pds} as dark histograms with error bars; the PDSs are log
scaled and grouped. It is clear that neither PDS can be described by a
simple power law, but both are well-fit by broken power laws,
indicating disk accretion \citep{vdk94,bok03,bko04}.  These PDS
features are not due to background fluctuations.  The PDSs of the {\it
XMM-Newton} background, shown with the light gray histograms in
Figure~\ref{pds}, do not show the variability seen in the PDSs of
r3-63.

We show in Figure~\ref{2004im} the 2004 July ACIS-I image of r3-63;
the circle indicates the 30$''$ extraction region used in our {\it
XMM-Newton} observations. There is no clear source for the observed
PDS within the 30$''$ apart from the SNR. Hence we conclude that the
accreting object is coincident with the SNR itself.

We modeled the EPIC-pn spectra of r3-63 from the 2002 and 2004
observations with XSPEC 11.3.1. All of the spectral fits from both
years were consistent.  Therefore we focus on the results from the
2002 January observation, which had the highest number of counts.

The 2002 January spectrum was grouped to ensure $>$20 count
bin$^{-1}$, and all counts outside the range 0.3--10 keV were
discarded, as the instrument response is not well known outside this
range. We found the useful range to be 0.3--2.1 keV.  Following K02,
we applied Raymond-Smith (RS) and non-equilibrium ionization (NEI)
plasma models, first fixing the abundance to Solar, then allowing it
to vary. We present the best fit models in Table~\ref{specmods}.
Fixed abundance values are provided in the table without any
associated errors.  Using previous knowledge about the SNR, we were
able to limit the number of free parameters and find evidence in the
spectrum for the presence of an LMXB in the SNR.

Like K02, the best-fitting model was an RS model with the O, Ne, and
Fe abundances free to vary.  All parameters of this fit are consistent
with the K02 results within the quoted error ranges.  However, this
fit has 6 free parameters in a 16 parameter model, and we could find
good fits by freeing almost any 6 of the parameters. The NEI fits with
fixed abundances had slightly better fits than the RS fits with fixed
abundances, but the NEI model has an additional free parameter in the
ionization timescale ($\log n_{\rm e}t$).

Most meaningful is the RS fit shown in the top panel of
Figure~\ref{xmmspec}, which had the abundances fixed to the values
measured from optical spectra (${{\rm N}\over{\rm H}} = 0.75 ({{\rm
N}\over {\rm H}})_{\odot}, {{\rm O}\over{\rm H}} = 0.27 ({{\rm O}\over
{\rm H}})_{\odot}$, and ${{\rm S}\over{\rm H}} = 0.44 ({{\rm S}\over
{\rm H}})_{\odot}$; \citealp{blair1982}).  This RS fit was much better
than the one that assumed solar abundances, with the same number of
free parameters.  Additionally, this fit yielded the same temperature
as K02, and like K02, this model fit had high $\chi^2/\nu$ (83/57).

The poor quality of the single-component spectral fits may be due to
the presence of a second component in the spectrum.  The spectrum of
r3-63 is dominated by the SNR, but all of the single-component fits to
the spectrum resulted in a slight hard excess, possibly due to a
second emission component.  The hard excess was also seen in the 2004
data.  When the 2004 spectra were fitted with a RS model with
abundances fixed to the values measured from optical spectra, the
parameters were consistent with those from the fit to the 2002
observation, and as shown in Figure~\ref{2004spec}, the fit resulted
in a hard excess.  In addition, the inability to achieve a good fit to
the 2002 data with the abundances fixed to the known values suggests
that another emission component may be contaminating the spectral
lines with continuum emission.

The spectral properties hint that the disk-accreting object detected
in the PDSs could be producing a hard-excess and enhancing the
continuum.  We therefore added a power-law component to the absorbed
RS and NEI models with the abundances fixed to the optical values. The
results are detailed in Table~\ref{2comp}; both cases show the
presence of a soft power-law component.  The NEI plus power-law model
has a slightly better fit again because of the extra free parameter.
The fit for the RS plus power-law model is shown in the bottom panel
of Figure~\ref{xmmspec}; there is no longer a hard excess in the
residuals to the fit.  This fit had $\chi^2/dof$ = 66/55, with the
power-law contributing 26$_{-16}^{+30}$\% of the X-ray luminosity.
According to an $F$-test, the power-law component is present with
0.998 significance.

The quality of the two-component fits was not as good as that of the
fit from the RS model with the abundances free to vary.  However, the
two-component models use the known abundances and our knowledge that
there is a variable source within the SNR, making these models more
meaningful in the context of our knowledge about r3-63.  The power-law
component is much softer than typical pulsars or HMXBs which have
photon indexes $\lap$1.  The results of the two-component spectral
fits are therefore more consistent with a power-law component coming
from the accretion disk of a low accretion rate LMXB, supporting what
is seen in the PDS.

The spectral data do not rule out a second component that is thermal
in nature.  We note that the hard excess can be equally well-fitted
with the addition of a blackbody component with kT$\sim$0.2 keV.  If
the second component is thermal, it could be from the SNR itself.
From the current data we cannot distinguish between the possibility
that the second component in the spectrum is from the SNR or from an
associated XRB.  Therefore, the spectrum does not provide conclusive
evidence for the presence of an XRB.  However, the variability
properties show that there is a variable source (i.e. an XRB)
contributing to the data, and both the {\it Chandra} (see
\S~\ref{chandra}) and {\it XMM-Newton} spectra of r3-63 are consistent
with the presence of an XRB.  These results suggest that the second
component in the spectrum is from the XRB.

\subsection{\it Chandra}~\label{chandra}

We examined the PDS and energy spectrum of the deepest available {\it
Chandra} observation of r3-63 to test their consistency with the {\it
XMM-Newton} results.  In addition we studied the 5 {\it Chandra}
observations that resolve the SNR into a shell.  The data were
sufficient to find a region within the SNR that is likely to be
variable.  Considering the results from the {\it XMM-Newton} data,
this region is likely the location of an associated X-ray binary.

We attempted to confirm the short term variability seen in the {\it
XMM-Newton} data using the deepest {\it Chandra} observation of the
region available in the archive.  This observation was the 37.7 ks
ACIS-S observation (Observation ID 1575).  While r3-63 was not
resolved in this observation, the 270 background-subtracted counts
from r3-63 could be used to test for variability.  However, the PDS
was featureless.  The 270 counts proved to be too few to detect the
variability seen with {\it XMM-Newton}.  This non-detection was not a
surprise, as the quality of the detection clearly depends strongly on
the number of counts in the observation.  This effect is noticeable in
the strength of the detections in the two {\it XMM-Newton} data sets.
The variability detection is stronger in the 2002 PDS with 2200 counts
than in the 2004 PDS with 700 counts (see Figure~\ref{pds}).

The second spectral component, which appears in the 2002 and 2004 {\it
XMM-Newton} spectra, is not inconsistent with the {\it Chandra}
spectral results from K02 or the lack of a variability detection in
the K02 {\it Chandra} data.  Such a component helps to explain the
poor spectral fits in K02 when the abundances measured from optical
spectra were applied.  In addition, the best-fitting model in K02 was
not consistent with the hardest spectral bin (see Figure 2 of K02),
hinting at the presence of a hard excess.  Finally, we refit the {\it
Chandra} spectrum (ObsID 1575) between 0.5--2.5 keV with a
two-component model, fixing the abundances, temperature, and photon
index to the values determined using the {\it XMM-Newton} data.  The
result was a good fit, with $\chi^2/\nu = 24.1/21$. This fit had only
2 free parameters: the normalization values of the 2 models.  A single
component RS model fit with the same fixed abundances but with $kT$
and $N_H$ free to vary (more free parameters than our 2 component fit)
over the same energy range does not fit the data as well, with
$\chi^2/\nu = 27.5/20$.  These fits are shown in Figure~\ref{cspec}.
The single-component fit clearly shows a hard excess that is
well-fitted by the addition of the power-law component, consistent
with the {\it XMM-Newton} results.  Furthermore, the two-component fit
to the {\it Chandra} spectrum indicates only 8\% of the flux came from
the second component.  Therefore only $\sim$20 of the counts in these
data are likely to be from the XRB, explaining the lack of a
variability detection in this observation.

\begin{figure}
\centerline{\psfig{file=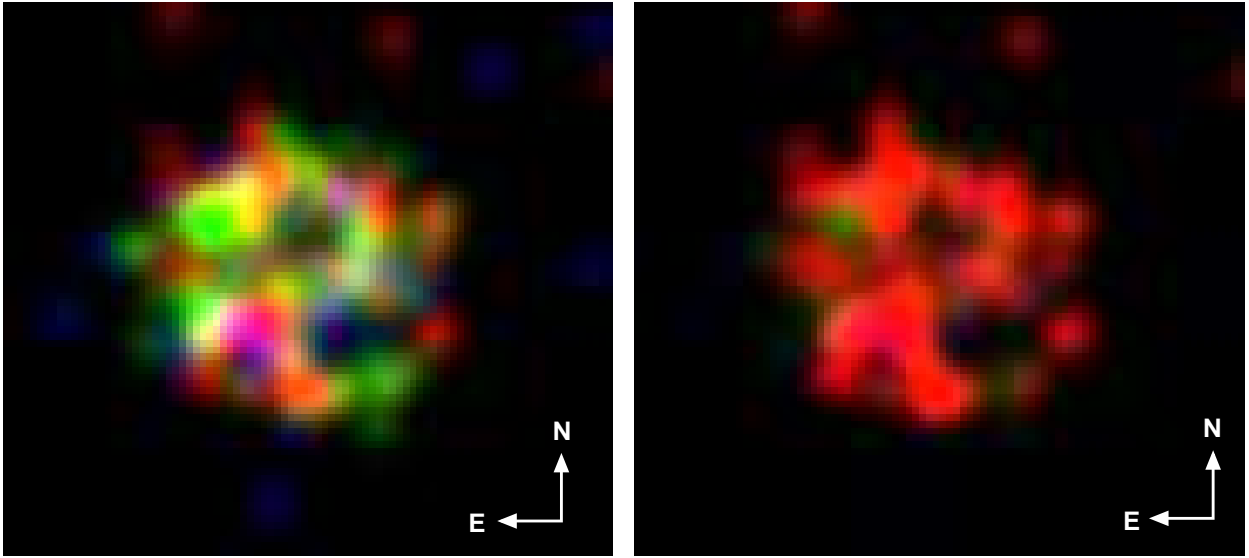,width=6.5in,angle=0}}
\caption{{\it Left:} Combined {\it Chandra} ACIS-I 0.3--7 keV image of
r3-63.  The colors represent different X-ray energy bands: 0.3--0.8
keV (red), 0.8--1.2 keV (green), and 1.2--7 keV (blue).  The image is
22$''$ across. The softest band was multiplied by 1.4 to correct for
decreasing sensitivity, but no correction for the {\it Chandra}
response was performed on this image.  {\it Right:} Same as left panel
with each band exposure-corrected to show the true relative fluxes.
This is the only resolved X-ray image of the SNR that fully corrects
for the {\it Chandra} response as well as all time-variable changes in
sensitivity.}
\label{im}
\end{figure}

\begin{figure}%}[!]
\centerline{\psfig{file=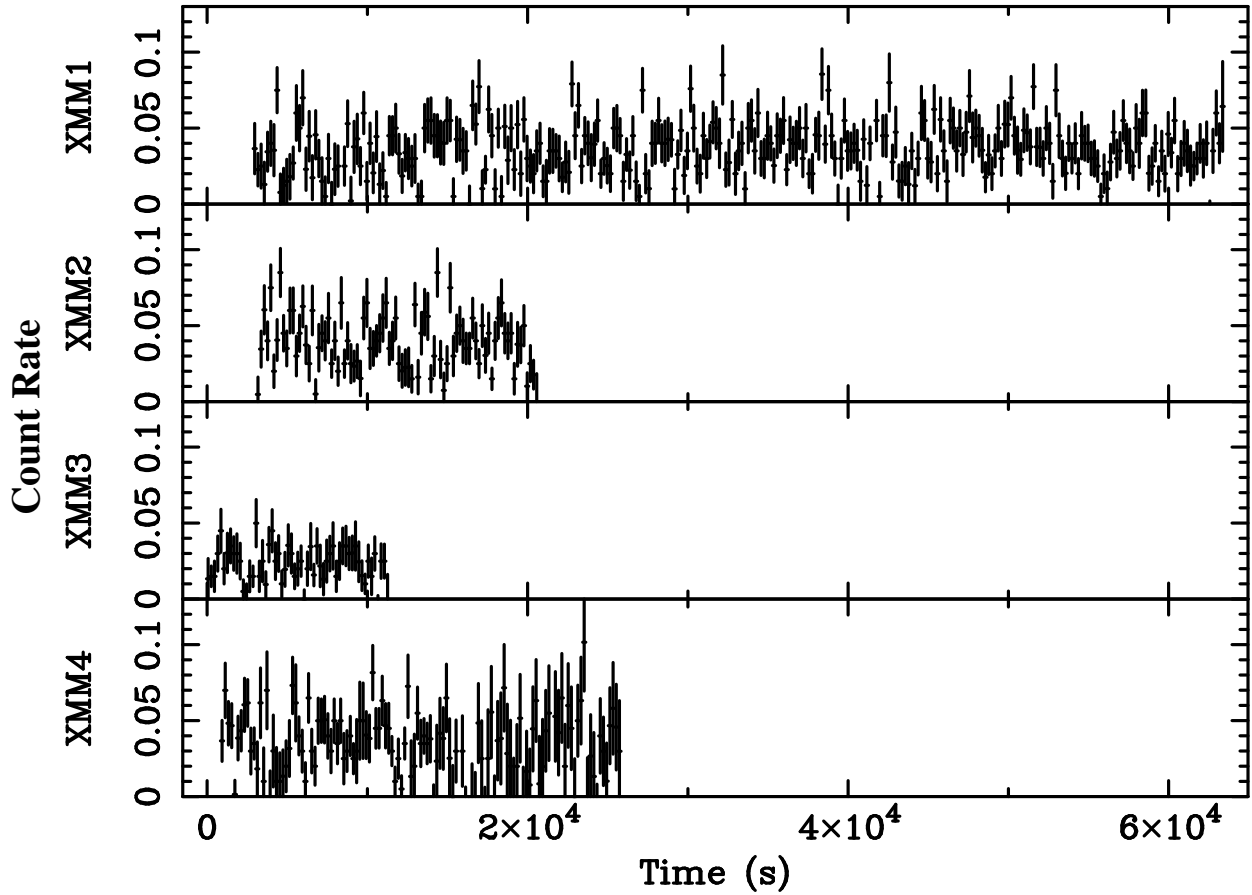,width=6.5in,angle=0}}
\caption{ Combined 0.3--2.0 keV {\it XMM-Newton}/EPIC lightcurves of r3-63
from the 2002 (top) and 2004 (bottom three) observations, binned to 200
s. The x- and y- axes are identical for all observations, to aid
comparison. }\label{lcs}
\end{figure}

\begin{figure}%*}[!t]
\centerline{\psfig{file=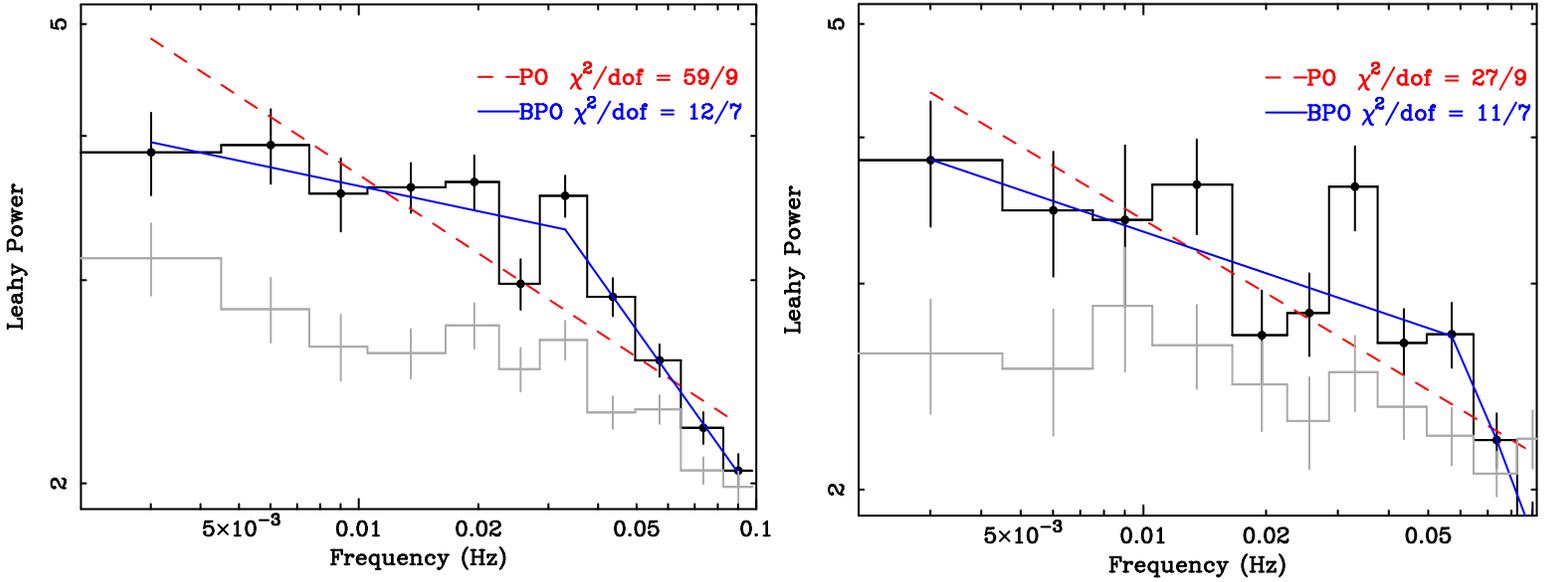,height=3in,angle=0}}
\caption{ Dark histograms with error bars show the power density
spectra (PDSs) of r3-63 from all of the 2002 {\it XMM-Newton} data
(left) and from the xmm2 (see Table~\ref{data}) 2004 data (right).
Light histograms with error bars show the PDSs of the background in
each observation.  The PDSs are log-scaled and Leahy-normalized so
that Poisson noise has a power of 2. The PDSs are averaged over many
333 s intervals with 5.2 s binning (64 bins per interval), and
grouped. Dashed lines show the best fitting single power-law function.
Solid lines show the best-fitting broken power-law function.  Both
PDSs are clearly described by broken power laws, indicating disk
accretion.  }
\label{pds}
\end{figure}

\begin{figure}%*}[!t]
\centerline{\psfig{file=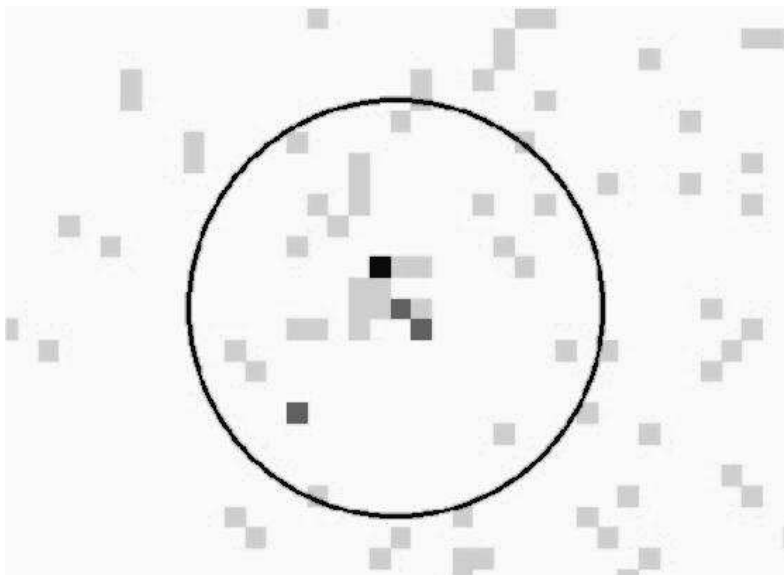,height=3in,angle=0}}
\caption{ACIS-I image of r3-63 from the 2004 July observation, taken
during the program of {\it XMM-Newton} observations. In this
observation r3-63 is too far off-axis to be resolved. The circle has a
radius of 30$''$, indicating the {\it XMM-Newton} extraction
region.}\label{2004im}
\end{figure}

\begin{figure}
\centerline{\psfig{file=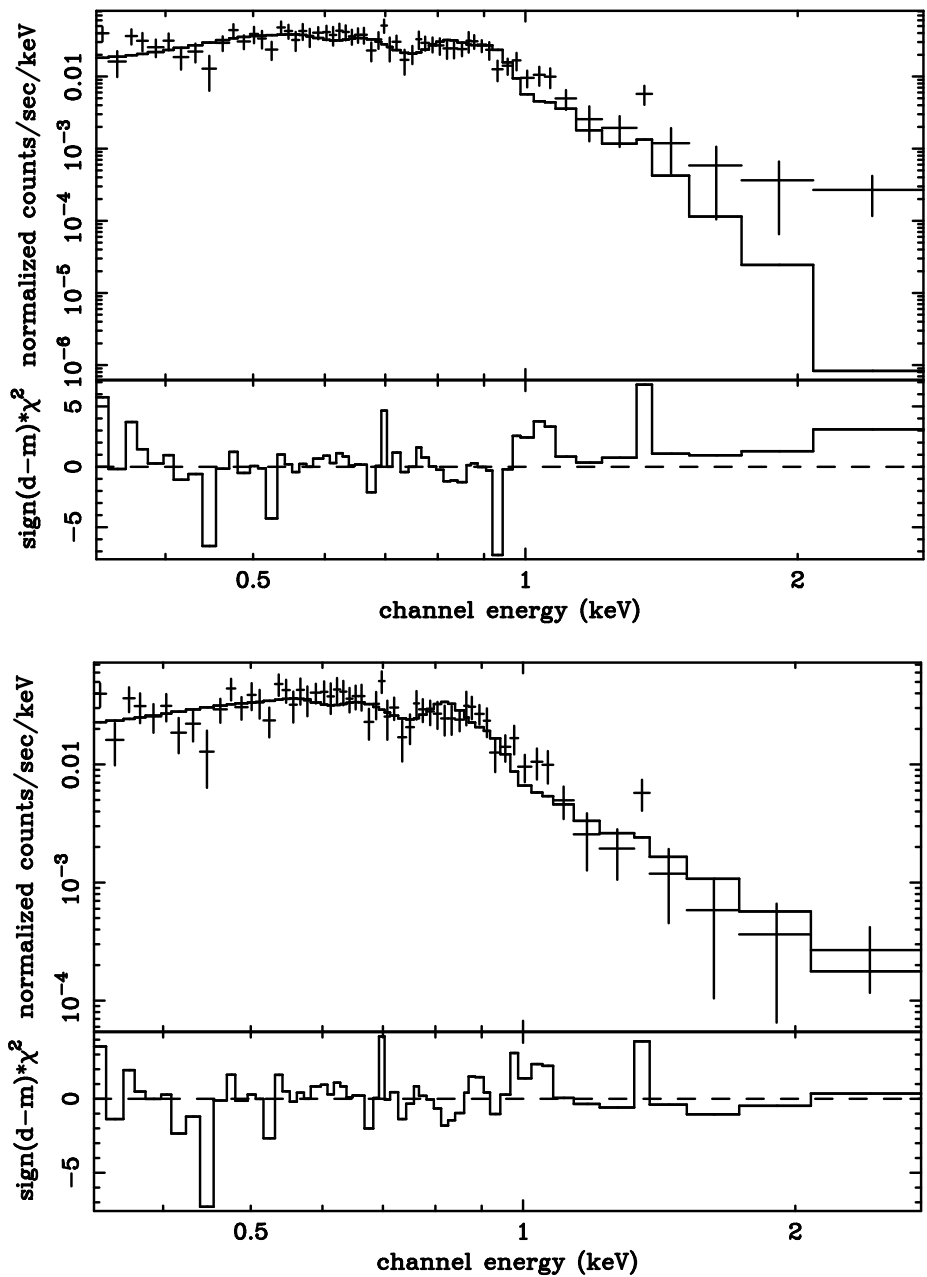,height=7in,angle=0}}
\caption{{\it Top:} Best fit Raymond-Smith model to the 0.3--2.1 keV
EPIC-pn spectrum from the 2002 observation of r3-63 with abundances
fixed to the values measured by \citet{blair1982}. The axes are
log-scaled, and the spectrum is folded.  The residuals to the fit are
plotted beneath the spectrum, showing a slight hard excess. {\it
Bottom:} Same as top, but with a power-law component added to the
model. The hard excess is no longer present.}
\label{xmmspec}
\end{figure}

\begin{figure}
\centerline{\psfig{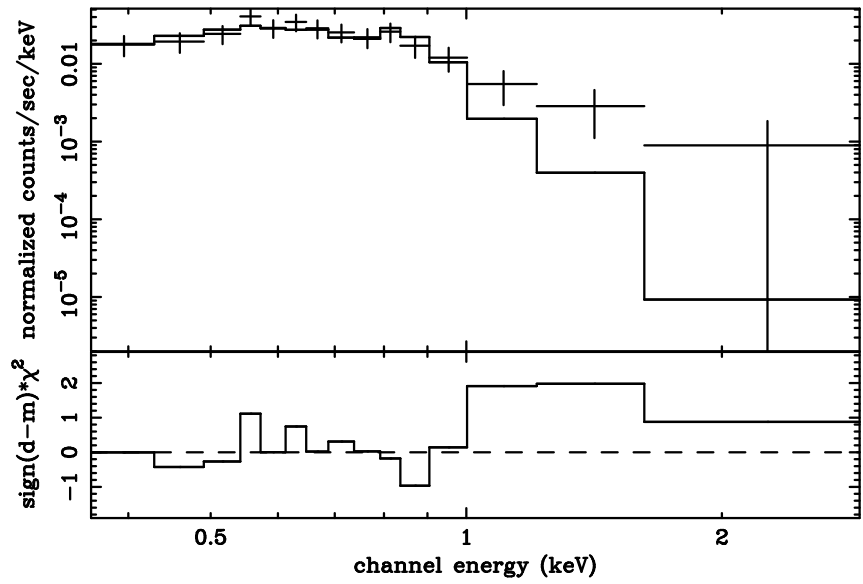}}
\caption{Best fit Raymond-Smith model to the 0.3--2.5 keV EPIC-pn
spectrum from the 2004 observation of r3-63 with abundances fixed to
the values measured by \citet{blair1982}. The axes are log-scaled, and
the spectrum is folded.  The residuals to the fit are plotted beneath
the spectrum, showing a hard excess consistent with that seen in the
single-component fit to the 2002 data.}
\label{2004spec}
\end{figure}

\begin{figure}
\centerline{\psfig{file=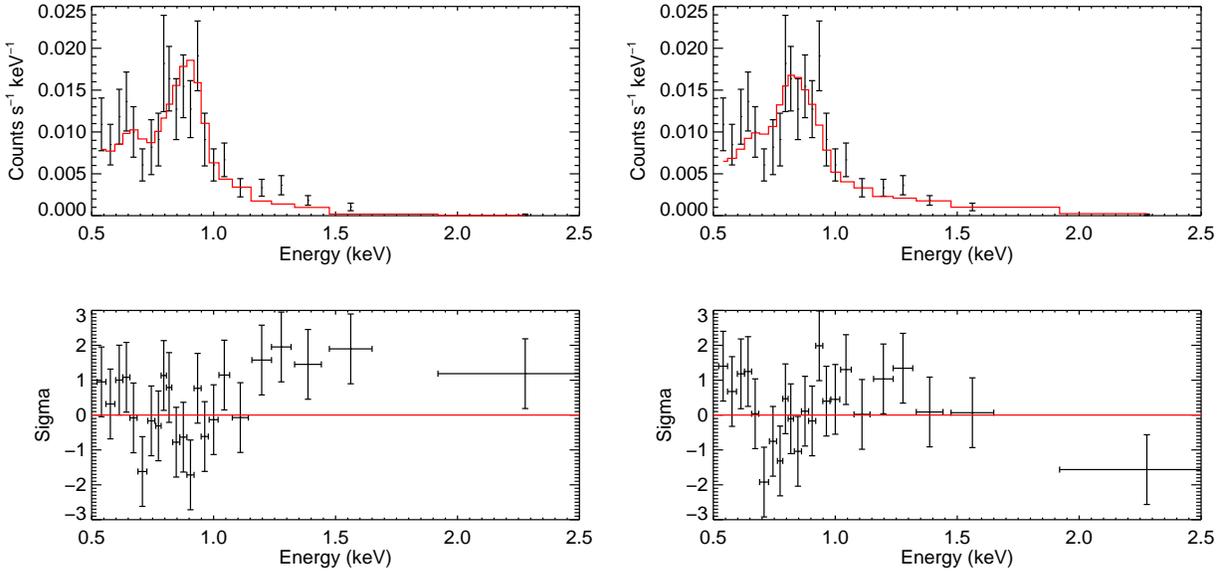,width=6.5in,angle=0}}
\caption{{\it Top left:} Best fit Raymond-Smith model to the 0.5--2.5
keV ACIS-S spectrum from the 37.7 ks observation of r3-63 with
abundances fixed to the values measured by \citet{blair1982}.  The
model is shown with the histogram, and the data are shown with the
error bars.  This fit had 3 free parameters: the normalization, the
temperature, and the absorption.  {\it Bottom left:} The residuals to
the fit are plotted beneath the spectrum, showing a slight hard
excess. {\it Top right:} Same as top left, but with a power-law
component added to the model. The abundances were fixed to the values
measured by \citet{blair1982}.  The absorption, temperature, and
photon index were fixed to the values measured from the EPIC-pn
spectrum, leaving the 2 normalization values as the only free
parameters.  {\it Bottom right:} The residuals to the fit are plotted
beneath the spectrum.  The hard excess is no longer present,
consistent with the {\it XMM-Newton} result.}
\label{cspec}
\end{figure}

Figure~\ref{im} shows the combination of all resolved {\it Chandra}
ACIS-I images of r3-63 we found in the data archive.  Direct
comparisons can be made between the left panel and K02 as the bands
chosen to represent each color are the same for both images; however,
it is important to note that the sensitivity of ACIS to soft ($\lap$1
keV) X-rays decreased by $\sim$15\% over the 9 month baseline of these
observations due to the build-up of contamination on the
OBF\footnote{http://cxc.harvard.edu/cal/Acis/Cal\_prods/qeDeg/index.html}.
Only the panel on the right of Figure~\ref{im} shows the image fully
corrected for the {\it Chandra} response, including the time-dependent
loss of sensitivity to soft X-rays.
 
The added depth in the left panel of Figure~\ref{im} reveals some
structure not easily seen in the 5 ks image from K02.  The overall
shell structure of the SNR is similar in both images, but the western
lobes of the shell are more apparent in the combined image. The added
depth shows fairly strong 0.8--1.2 keV emission in the western half of
the SNR.  In addition, the K02 image appeared strongly red (0.3--0.8
keV) on the north rim of the shell, the combined image has only small
pockets of red on the north rim.  The rest of the SNR appears to
exhibit very similar colors to the K02 image, and the soft-band
portion of our deeper image was corrected for decreased
sensitivity. Therefore the greener north rim appears attributable to
the increase in depth rather than to the decrease in sensitivity to
soft ($<$1 keV) X-rays over the course of the observations.

The exposure-corrected image (right panel of Figure~\ref{im}) shows
the true relative X-ray fluxes of the different bands.  It is clearly
dominated by soft emission, consistent with the SNR spectrum.  

\begin{figure}
\centerline{\psfig{file=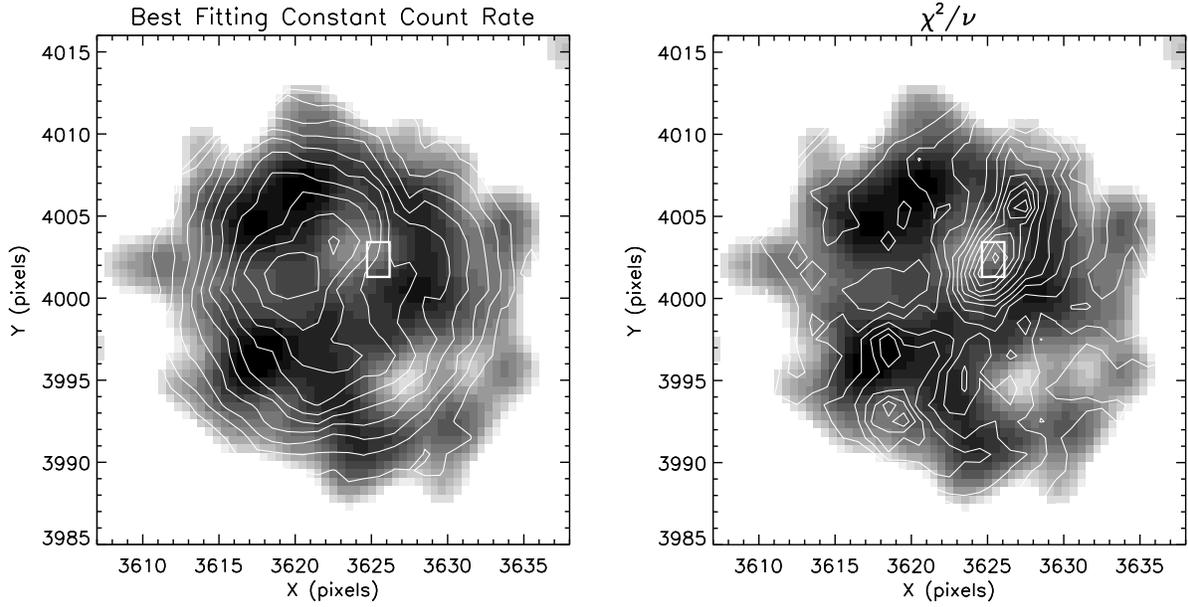,width=6.5in,angle=0}}
\caption{{\it Left:} Contours of the count rate as a function of
aperture center position from the $\chi^2$ lightcurve analysis are
overplotted on the X-ray image of the SNR.  Contours are spaced at
intervals of 0.0002 ct s$^{-1}$ and cover the range 0.0006--0.0024 ct
s$^{-1}$.  A white box marks the location of the center of all
apertures with mean count rates $>$0.0015 ct s$^{-1}$ and
$\chi^2_{\nu} >$1.945. {\it Right:} Contours of the best-fitting
$\chi^2_{\nu}$ value as a function of aperture center position from
the $\chi^2$ lightcurve analysis are overplotted on the same X-ray
image. Contours are evenly spaced at intervals of 0.2 and cover the
range 0$\leq\chi^2_{\nu}\leq2.2$.  Notice the area of high
$\chi^2_{\nu}$ in the northwest quadrant of the SNR.  The white box
marks the same location as in the left panel.}
\label{chi2grey}
\end{figure}

More intriguing are the results of the variability analysis.  The SNR
sampled as a whole was well-fitted ($\chi^2_{\nu}=0.8$) by a constant
count rate of 0.0067 ct s$^{-1}$; however when the SNR was broken into
smaller regions, the lightcurves of regions northwest of the center
showed evidence for variability.  These results are shown in the right
panel of Figure~\ref{chi2grey} with contours of the $\chi^2_{\nu}$
values for fits to a constant count rate overplotted on the greyscale
{\it Chandra} X-ray image of the SNR.  There is a noticeable area of
high $\chi^2_{\nu}$ values in the northwest quadrant of the SNR,
suggesting that if any portion of the SNR varied over the course of
the observations, it was this portion.   

For comparison, contours of the best-fitting constant rates are shown
overplotted on the same greyscale image in the left panel of
Figure~\ref{chi2grey}.  The contours in this panel show the effects of
our aperture size, which effectively re-samples the data to 6$''$
spatial resolution resulting in the loss of most of the structure seen
at full resolution.  The only detail that survives the re-sampling is
the off-center peak in emission indicating that the eastern half of
the SNR is brighter than the western half.

We determined the threshold for variability detection in our low-count
lightcurves.  With the 4 degrees of freedom offered by our 5-point
lightcurves, a $\chi^2_{\nu}$ value of 1.945 corresponds to a 90\%
probability for variability in the lightcurve.  We ran Monte Carlo
simulations drawing 5 random lightcurve points with 3--19 counts in
each.  This range of counts represented the range seen in the
lightcurves from our data.  We then applied Poisson errors and
computed $\chi^2_{\nu}$ in the same fashion as applied to our data.
In 10$^4$ simulations, 87\% of the lightcurves had
$\chi^2_{\nu}<1.945$, similar to the prediction of 90\% from $\chi^2$
statistics.  

We also created a lightcurve of the background in our combined r3-63
image with an extraction area of 0.34 arcmin$^2$ so that the number of
counts per bin were $\sim$10.  Poisson errors were used.  This
lightcurve had $\chi^2_{\nu}=0.38$ when fitted to a constant rate,
clearly indicating that the background was not responsible for any
variability seen in the r3-63 lightcurves.  In addition, the mean
background count rate in a 3$''$ radius aperture was just 2\% of the
rate observed in apertures of identical size in r3-63, further
eliminating the background as a source of variability in our
lightcurves.  The background lightcurve is shown with that of the
variable region in Figure~\ref{apim}.

We determined the location of the most variable region in the SNR.
The mean position of all aperture centers with $\chi^2_{\nu} > 1.945$
and best-fit count rates $>$0.0015 ct s$^{-1}$ was $X=3625.2\pm0.6,
Y=4002.2\pm0.9$.  This location is marked with a white box on
Figure~\ref{chi2grey}.  The corresponding R.A. and DEC. are
00:43:27.81$\pm$0.3$''$ (0.030$^s$), 41:18:31.2$\pm$0.4$''$
(J2000). The location of this centroid shows that the elongation of
the variable region toward the outer portion of the SNR (northwest of
the centroid) is due to very low count rates in those outermost
apertures.

\begin{figure}
\centerline{\psfig{file=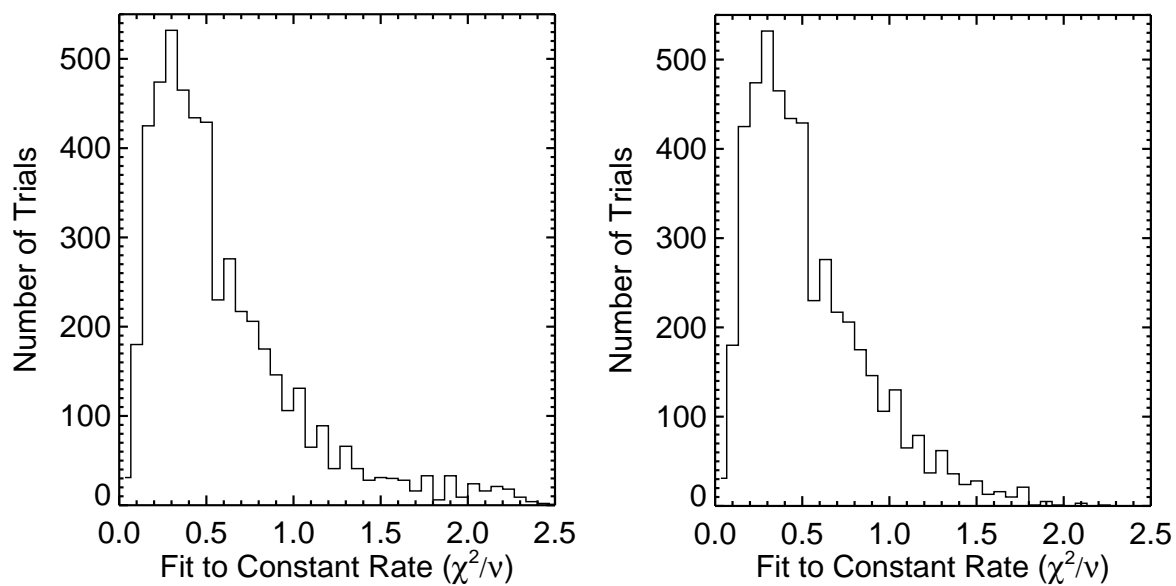,width=6.5in,angle=0}}
\caption{{\it Left:} Histogram the $\chi^2_{\nu}$ value vs. the number
of random apertures with 6 pixel radius that result in a lightcurve
with that value when fit to a constant count rate.  All apertures with
mean count rates $>$0.0015 count s$^{-1}$ are included.  {\it Right:}
Same as left panel with the results from all apertures with centers at
$X=3625.5\pm1.4, Y=4002.5\pm1.8$, as well as those with mean count
rates $<$0.0015 count s$^{-1}$, removed.}
\label{chihist}
\end{figure}

\begin{figure}
\centerline{\psfig{file=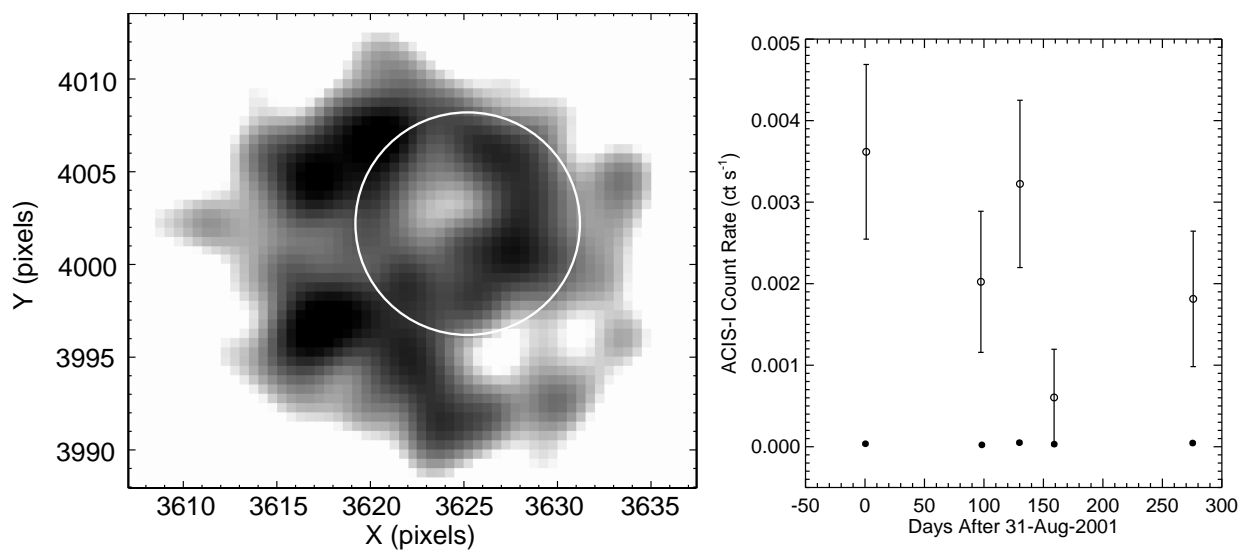,width=6.5in,angle=0}}
\caption{{\it Left:} Smoothed 0.3--7 keV image of r3-63.  Darker areas
denote higher X-ray flux.  The white circle shows the 6 pixel aperture
centered on the most variable portion of the SNR. {\it Right:} The
lightcurve of the aperture shown is marked with the open circles.
This lightcurve has a best-fitting constant rate of 0.0018 ct
s$^{-1}$, and $\chi^2/{\nu}$=9/4, which has 94\% probability of
variability.  Filled circles mark the lightcurve for the background
scaled to an equivalent area.  The points are larger than the errors
in the background count rates.}
\label{apim}
\end{figure}

We investigated the impact of the most variable region on the
$\chi^2_{\nu}$ distribution.  The left panel of Figure~\ref{chihist}
shows a histogram of the $\chi^2_{\nu}$ values measured for the
lightcurves of all aperture centers with count rates $>$0.0015.  The
distribution falls off quite smoothly to $\chi^2_{\nu}$=1.5; then
there appears to be some excess trials with $\chi^2_{\nu}>1.5$.  The
right panel shows the same histogram with all aperture centers in the
region $X=3625.2\pm1.2, Y=4002.2\pm1.8$ removed.  The excess at high
$\chi^2_{\nu}$ disappears, providing further assurance that this
region of the SNR is the most likely to be variable.

An aperture at the most variable position in the SNR and the
corresponding lightcurve are shown in Figure~\ref{apim}. The
lightcurve has $\chi^2/{\nu}$=9/4, which is a detection of variability
with 94\% confidence according to standard $\chi^2$ statistics.  We
checked this reliability with Monte Carlo simulations performed as
described above.  In 10$^4$ simulations, 94\% had $\chi^2/{\nu}<$9/4,
as expected from $\chi^2$ statistics.

The poor fits to a constant count rate seen in the northwest portion
of the SNR do not provide a highly reliable detection of variability,
as the high $\chi^2/{\nu}$ value is clearly due to only one outlying
data point with a low count rate.  Furthermore, in order to isolate
this variability in the {\it Chandra} data, we have eliminated about
75\% of the SNR flux\footnote{The SNR total count rate is 0.0067 count
s$^{-1}$, and the mean count rate of the variable region is 0.0018
count s$^{-1}$}.  On the other hand, considering the {\it XMM-Newton}
data, which show that the SNR contains an XRB, this region is the most
likely location of an XRB that can be found with currently-available
data.  Further on-axis {\it Chandra} observations will be required to
ascertain this possible detection.

Additional concern regarding the variability of this region of the SNR
is warranted because it contains some soft (0.3--0.8 keV) emission and
the lightcurve shows a decrease in flux over the 9-month baseline of
the observations.  The build-up of contamination on the OBF surely
contributed somewhat to this decrease in count rate (see
\S~\ref{datsec}); however, the level of variability observed cannot be
explained by OBF contamination effects.  The effect of the
contamination was $\sim$40\% loss of sensitivity at 0.5 keV from April
2000 to June
2002\footnote{http://cxc.harvard.edu/cal/Acis/Cal\_prods/qeDeg/index.html}.
Therefore, assuming the decrease was linear with time, during the 9
months between the first and last resolved observation of r3-63, the
sensitivity dropped by $\sim$15\% at 0.5 keV.  We tested the effects
of this decrease in sensitivity by artificially lowering the count
rates of our Monte Carlo simulations by 0--20\%, increasing by 4\% at
each consecutive artificial data point.  The simulations still showed
at least 94\% of the artificial lightcurves had $\chi^2/{\nu}<$9/4
when fitted to a constant rate.  The variability does not appear to be
due to the effects of the OBF contamination.

The SNR center is $X=3623.5\pm1.0, Y=3999.9\pm1.0$
(00:43:27.88$\pm$0.5$''$ (0.04$^s$), 41:18:30.1$\pm$0.5$''$).  The
most variable position in the SNR is therefore 0.8$'' \pm0.6''$ west
and 1.1$'' \pm0.7''$ north of the SNR center, 1.4$''\pm0.9''$ distant
from the SNR center.  This location provides additional evidence that
the XRB is associated with the SNR, and it constrains the velocity of
the XRB with respect to the SNR center.

The probable location of the XRB makes it likely to be associated with
the SNR.  The mean density of point sources with 0.3--7 keV
luminosities $>2\times10^{35}$ erg s$^{-1}$ is 0.4 sources
arcmin$^{-2}$ in the area from 7.5$'$ to 8.5$'$ from the center of
M31, according to a search of the \citet{kong2002acis} catalog.  Using
this value as the density of sources at the galactocentric distance of
r3-63 (8.5$'$), the probability of an unassociated X-ray source
falling inside the 5.5$''$ radius of the SNR is 1\%.  The probability
decreases to 0.1\% for an unassociated source to fall within 2$''$ of
the SNR center.

Because an XRB at the variable position in the SNR was likely created
by the SN event, we can use the distance of the XRB from the SNR
center along with the SNR age to constrain the projected velocity of
the XRB.  Assuming a distance of 780 kpc to M31
\citep{williams2003sfh}, this separation corresponds to a projected
distance of 5.3$\pm$3.4 pc.  Taking the SNR age to be 21$^{+1}_{-5}$
kyr, from the best-fitting spectral parameters of K02, the projected
velocity of the binary system with respect to the SNR center is
240$^{+280}_{-160}$ km s$^{-1}$.

\section{Discussion: HMXB or LMXB?}

There is no foolproof method for determining whether the variable
source in r3-63 is a HMXB or LMXB from the currently-available data.
We therefore use the clues available to make an indirect argument that
the source is a LMXB.  This indirect argument is based on available
optical images, X-ray images, spectra, PDSs, and lightcurves.  We then
investigate the possibility of the existence of a SNR/LMXB association
in r3-63 based on the properties of LMXBs and current models of LMXB
formation from SNe.

\begin{figure}
\centerline{\psfig{file=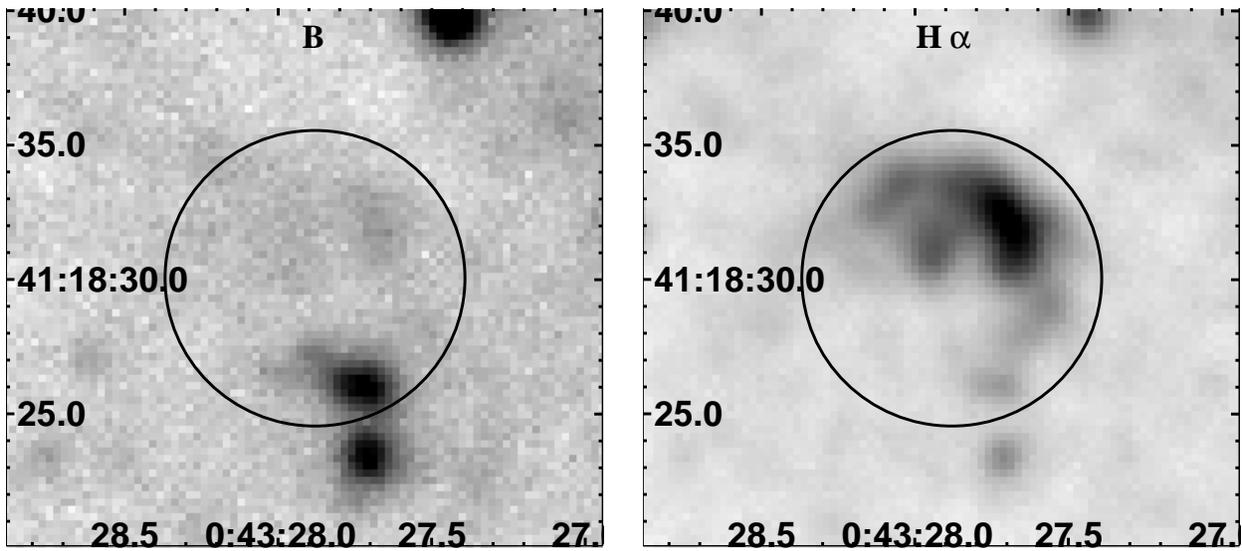,width=6.5in,angle=0}}
\caption{{\it Left:} The Local Group Survey $B$-band image of r3-63.
The image shows the same patch of sky shown by the X-ray image in
Figure~\ref{im} down to $B\sim21$.  The black circle shows the 11$''$
diameter of the SNR shell, revealing one bright star ($B\sim19.4$) in
the shell 5$''$ southwest of the SNR center. {\it Right:} The Local
Group Survey H$\alpha$ image of r3-63 is shown for comparison to the
broadband image so that the relative positions of continuum sources
and the SNR shell can be compared.}
\label{lgs}
\end{figure}

The optical counterpart search of \citet{williams2004hrc} using the
data of the Local Group Survey (LGS; \citealp{massey2001}) revealed no
stellar counterpart to r3-63 down to $V=20.6$.  Broadening the search
of these data to include the full diameter of the SNR (see
Figure~\ref{lgs}, left panel) reveals a bright star within the shell
region. The circles on the LGS $B$-band and H$\alpha$ images in
Figure~\ref{lgs} mark the 11$''$ diameter shell of the SNR. The
H$\alpha$ image is shown here only to facilitate comparisons between
the locations of the optical shell and continuum sources in the region.
There is a $B$=19.4 (M$_B\sim-5.4$ at distance and typical extinction
to M31) star near the southwest edge of the shell.  This star is 5$''$
(19 pc) from the SNR center.  If this star is a HMXB, it is unlikely
associated with the SNR, as any kick that would move the binary 19 pc
in 20 kyr (900 km s$^{-1}$) would disrupt the binary.

No bright star ($B<21$) is seen in the northwest portion of the SNR.
Therefore, if the X-ray variability seen in the {\it Chandra} data is
real, then the XRB that caused it is not a bright HMXB with a
companion earlier than B1 (M$_B<-4$).  On the other hand, the optical
data still allow an HMXB with a later B star secondary, and if the
probable variability seen in the {\it Chandra} data is not real, there
could be an {\it unassociated} bright HMXB in the SNR.  Since there
are few known HMXBs in M31 \citep{williams2004hrc}, such a coincidence
is unlikely.  We therefore conclude that the available optical data do
not yield evidence to support the idea that the XRB in r3-63 is an
HMXB, but they do not completely rule out the possibility.

In addition, most young HMXBs are Be/X-ray systems that contain
pulsars \citep{liu2000}.  These HMXBs typically have hard X-ray
spectra.  The power-law component of the spectrum of r3-63 is soft.
Furthermore, the PDSs of pulsars are dominated by pulsations. The lack
of a peak in the PDS of r3-63 (see Figure~\ref{pds}) indicates the
variability is not dominated by pulsations.  These data do not rule
out the possibility that a pulsar could be present, but they provide
no evidence for the presence of pulsations or a HMXB.  We therefore
suggest the SNR contains an associated LMXB because of (1) the
appearance of the region in optical images, (2) the appearance of the
{\it XMM-Newton} extraction region in the resolved contemporaneous
{\it Chandra} data, (3) the variability seen in the {\it Chandra}
resolved lightcurves, (4) the softness of the X-ray spectrum, and (5)
the properties of the {\it XMM-Newton} PDS.
 
A LMXB in r3-63 is reasonable considering the properties of known LMXB
systems and models of LMXB formation and evolution.  The probable
position of the XRB is consistent with current systemic and kick
velocity distributions, and LMXB formation models allow for the
possibility that X-ray activity could begin in an LMXB shortly after
the SN event.

The more central portion of the variable region in r3-63 (discussed in
\S~3.2) provides the most reasonable velocity of the binary away from
the SNR center ($\sim$100 km s$^{-1}$).  Several Galactic black hole
binaries have projected orbital velocities of $>$100 km s$^{-1}$ (see
values in \citealp{orosz2003}), suggesting binary systems can survive
such a kick velocity.  In addition, \citet{kalogera1998} show that
LMXBs can survive in simulations with mean kick velocities of 100--200
km s$^{-1}$, and \citet{fryer1998} show that for a double-peaked kick
velocity distribution, some binaries survive kicks in the low velocity
peak.  Such a double-peak is seen in the velocity distribution of
pulsars, with a low velocity peak at 90 km s$^{-1}$
\citep{arzoumanian2002}, and kicks from asymmetric SNe could produce
such velocities \citep{scheck2004,fryer2004}.  Finally, the low end of
our allowed velocity range for the XRB in r3-63 overlaps the results
of earlier simulations of the systemic velocities of binary systems
that survive a SN event to become LMXBs (180$\pm$80 km s$^{-1}$,
\citealp{brandt1995}).

A SNR/LMXB association would be the first of its kind, but such
objects may be extremely rare because SNRs fade quickly and a
surviving binary could take some time to become X-ray active.  From a
theoretical point of view, both HMXBs and LMXBs form from SNe, and
since many examples of both types of systems are known in our Galaxy,
some of these systems clearly survive the SN event.  

Survival of a SN is easier if the secondary star is massive. In
spherically symmetric explosions, i.e. those without a kick, the
binary will become unbound if more than 50\% of the pre-SN mass is
ejected. This 50\% criterion is more likely to be met in low mass
systems.  If the SN provides a kick, this rule no longer holds, but
for small to intermediate kick velocities the survival probability
still increases with companion mass \citep{kalogera1996}.

While it is true that the parameter space for the formation of neutron
star LMXBs is very small \citep{king1997,kalogera1998,kw1998}, many
LMXBs must survive their birth SNe to create the known population.  In
addition, many LMXBs are transient sources, suggesting that there may
be a large, dormant, as yet undiscovered population which would add to
the SN survival rate.  This possibility that many low-mass binaries
survive SNe is confirmed by some evolutionary calculations, in which
LMXBs form for a wide range of kick velocities
(e.g. \citealp{brandt1995,kalogera1998}).

After the SN event, most systems must evolve into Roche-lobe contact
before mass transfer, and hence X-ray activity, starts.  For the onset
of mass-transfer to occur within $\sim$20 kyr of the SN requires the
system to have emerged from the SN almost semi-detached, while
avoiding a merger in the immediate eccentric post-SN orbit.  The
potential for low-mass systems not to evolve into LMXBs until after
the SNR fades, coupled with the fact that the systemic velocities of
HMXBs are on average smaller than those of LMXBs, would favor the
appearance of a HMXB inside its parent SNR over that of a LMXB.

Even though HMXBs may be more likely to survive a SN and begin active
accretion while their parent remnant is still bright, very few
SNR/HMXB associations are known (see \S~1).  If the conditions
necessary to produce a SNR/LMXB association are even less likely, as
suggested by the formation simulations and timescale arguments,
SNR/LMXB associations would certainly be rare, but they could exist.

\section{Conclusions}

We have studied in detail the timing analysis of the X-ray emission
from the M31 SNR r3-63 using both {\it XMM-Newton} and {\it Chandra}.
The {\it XMM-Newton} data provide PDSs which show variability within
the SNR lightcurve at very high confidence.  We can associate Type A
variability with the SNR, indicating disk accretion and probable Roche
lobe overflow.  The accreting object is most likely a neutron star or
black hole in a binary system.  These data therefore contain the
signature of an accreting X-ray binary very close to the SNR,
presumably formed by the SN event.  If Roche lobe overflow is
responsible for the mass transfer, the binary is likely a LMXB, which
would make r3-63 the first SNR known that contains a LMXB.

Although deeper resolved optical and X-ray data are required to
confirm the lack of a HMXB near the center of r3-63, we see no clear
evidence for a HMXB in the currently-available data. On the other
hand, the data are all consistent with an associated LMXB in r3-63.
If we have found the first SNR/LMXB association, we have had to search
outside the Galaxy within the largest spiral in the Local Group to
find it.  Confirmation of the associated LMXB would support the idea
that such associations are rare, but would also show that LMXBs can
start mass-transfer very shortly after the formation of the compact
object.

Detailed analysis of the spatially-resolved lightcurve from all
resolved detections of r3-63 in the {\it Chandra} archive suggest
that, while the SNR taken as a whole is not significantly variable,
there is variability detected at the 94\% confidence level
1.4$''\pm$0.9$''$ northwest of the SNR center which cannot be
attributed entirely to ACIS contamination issues or photometric
errors. Further observations are needed to confirm the variability,
determine a more precise SNR age, and directly measure the position of
the XRB. 

The current age estimate of the SNR and XRB, along with the possible
XRB position, provide an estimate of the projected systemic velocity
away from the SNR center.  If the variability seen in the {\it
Chandra} data is real and marks the location of the XRB detected by
{\it XMM-Newton}, then the XRB associated with the SNR is 5.3$\pm$3.4
pc (projected distance) from the SNR center.  With a SNR age of
21$^{+1}_{-5}$ kyr (K02), this separation corresponds to an XRB
velocity estimate of 240$^{+280}_{-160}$ km s$^{-1}$ with respect to
the SNR center.  The low end of this range is reasonable given current
knowledge of the distribution of neutron star and black hole
velocities \citep{arzoumanian2002,orosz2003}, the expected kick
velocities from asymmetric SNe \citep{scheck2004,fryer2004}, and the
results of simulations of binary survival of SN kicks
\citep{brandt1995,kalogera1998}.  On the other hand, the system could
not have remained bound and acquired a systemic velocity close to the
high end of this range.

Presently, r3-63 appears to be an excellent candidate SNR/LMXB
system. The broken power law PDS, the soft X-ray spectrum, the
non-detection of pulsations, and the lack of bright stars near the
probable XRB position all support the idea that r3-63 contains a
LMXB. Such systems must be very rare, possibly due to the supernova
survival rate of low mass binary systems and/or the timing necessary
for a low-mass system to begin mass-transfer while its birth SNR is
still bright. More direct evidence is needed to confirm this
association and determine if SNR/LMXB systems exist.

Support for this work was provided by NASA through grant number
GO-9087 from the Space Telescope Science Institute and through grant
number GO-3103X from the {\it Chandra} X-Ray Center.  MRG acknowledges
support from NASA LTSA grant NAG5-10889.  RB was funded by PPARC.

%\bibliography{apjmnemonic,references} 
%\bibliographystyle{apj}

\clearpage

\begin{deluxetable}{ccccccccccc}
%\tablewidth{in}
\tablecaption{Observations used for the study.}
\tableheadfrac{0.01}
\tablehead{
\colhead{{ObsID}} &
\colhead{{Date}} &
\colhead{{R.A. (J2000)}} &
\colhead{{Dec. (J2000)}} &
\colhead{{Roll (deg.)}} &
\colhead{{Exp. (ks)}} &
\colhead{{Sep. ($'$)}}
}
\startdata
1577 & 2001 Aug 31 & 00 43 07.3 & 41 19 17.7 & 143.0 & 4.9 & 4.0\\
2895 & 2001 Dec 07 & 00 43 05.4 & 41 17 33.6 & 271.7 & 4.9 & 4.4\\
2897 & 2002 Jan 08 & 00 43 09.9 & 41 18 43.8 & 292.4 & 4.9 & 3.4\\
2896 & 2002 Feb 06 & 00 43 06.0 & 41 16 47.2 & 309.5 & 4.9 & 4.4\\
2898 & 2002 Jun 02 & 00 43 10.5 & 41 19 16.4 & 88.6 & 4.9 & 3.3\\
xmm1 & 2002 Jan 06 & 00 42 39.1 & 41 15 47.0 & 250 & 64.5 & 9.8 \\
xmm2 & 2004 Jul 16 & 00 42 42.1 & 41 16 57.2 & 64 &  20.3 & 8.7\\
xmm3 & 2004 Jul 18 & 00 42 42.3 & 41 16 58.2 & 63 &  21.9 & 8.7\\
xmm4 & 2004 Jul 19 & 00 42 42.2 & 41 16 57.2 & 63 &  27.1 & 8.7\\
\enddata
\label{data}
\end{deluxetable}

\begin{table*}[!t]
\centering
\renewcommand{\tabcolsep}{0.5pt}
\caption{Results of fitting Raymond-Smith and non-equilibrium ionization plasma
models to the 2002 January 6 XMM-Newton spectrum of r3-63. Uncertainties are
quoted at a 90\% confidence level for fits with $\chi^2$/dof
$<$2.  Fixed abundance values are given with no uncertainties.}
\label{specmods}
\begin{tabular}{lcccccccccccccc}
\noalign{\smallskip}
\hline
\noalign{\smallskip}

 & $N_{\rm H}$ & k$T$  & $\log$ $n_{\rm e}t$ & N & O & Ne & S & Fe &
$L_{0.3-10}$ & \\
 Model & ($\times 10^{21}$ cm$^{-2}$) & (keV) & &  &&&&&($\times$10$^{37}$ erg
s$^{-1}$) & $\chi^2$/dof\\
\noalign{\smallskip}
\hline
\noalign{\smallskip}
 RS &  0.6 & 0.25 & \nodata  & 1 & 1 & 1 & 1& 1 &   0.7 & 122/57\\
   & 4.1$^{+0.6}_{-0.7}$ & 0.143$^{+0.008}_{-0.004}$ & \nodata & 0.75 & 0.27 &
1 & 0.44 & 1 & 1.52$^{+0.14}_{-0.5}$ & 83/57\\
  & 0.7$^{+1.8}_{-0.6}$ & 0.31$^{+0.06}_{-0.08}$ & \nodata  & 1 &
0.33$^{+0.19}_{-0.12}$ & 0.51$\pm$0.22 & 1 & 0.13$^{+0.08}_{-0.05}$ &
0.90$\pm$0.14  & 58/54\\
NEI & 1.0$^{+1.1}_{-0.7}$ & 1.3$\pm$0.7 & 10.1$\pm$0.2 & 1& 1& 1 &1 &1 &
0.8$^{+0.2}_{-0.5}$ & 74/56\\
 & 2.6$\pm$0.8 & 0.186$\pm$0.012 & 13.5$_{-0.3}^{+0.2}$ & 0.75 & 0.27 & 1 &
0.44 & 1 & 4.0$\pm$? & 77/56\\

\noalign{\smallskip}
\hline
\noalign{\smallskip}
\end{tabular}
\end{table*}

\begin{table*}[!t]
\centering
\renewcommand{\tabcolsep}{0.5pt}
\caption{Results of fitting Raymond-Smith and non-equilibrium
ionization plasma models plus a power-law component to the 2002
January 6 XMM-Newton spectrum of r3-63. Abundances were fixed to
values measured from \citet{blair1982}. Uncertainties are quoted at a
90\% confidence level for fits with $\chi^2$/dof $<$2.}
\label{2comp}
\begin{tabular}{lcccccccccccccc}
\noalign{\smallskip}
\hline
\noalign{\smallskip}

 & $N_{\rm H}$ & k$T$  & $\log$ $n_{\rm e}t$ & photon &
$L_{0.3-10}$ (total) & $L_{0.3-10}$ (power-law) &\\
 Model & ($\times 10^{21}$ cm$^{-2}$) & (keV) & & index &($\times$10$^{37}$ erg s$^{-1}$) & ($\times$10$^{37}$ erg s$^{-1}$) & $\chi^2$/dof\\
\noalign{\smallskip}
\hline
\noalign{\smallskip}
 RS  & 2.8$^{+1.3}_{-0.9}$ & $0.16\pm0.02$ & \nodata & $3.2\pm1.5$ & 3.1$^{+0.9}_{-0.4}$ & $0.8_{-0.5}^{+0.9}$ & 66/55\\
NEI & 2.3$^{+0.7}_{-0.6}$ & 0.189$^{+0.014}_{-0.013}$ & 13.69$_{-0.18}^{+0.01}$ & 3.7$^{+1.3}_{-1.6}$ & 2.5$^{+0.8}_{-0.3}$ & 0.07$^{+1.2}_{-0.07}$ & 61/54\\

\noalign{\smallskip}
\hline
\noalign{\smallskip}
\end{tabular}
\end{table*}

\clearpage

\clearpage
%*}

%*}

\end{document}